\documentclass[12pt]{article}
\usepackage[dvips]{graphicx}
\usepackage{pdproc}

  \makeatletter 
  \def\@cite#1{[#1]} 
  \makeatother    
\begin{document}

\renewcommand{\thefootnote}{\alph{footnote}}

\title{ Raising the Higgs Mass in Supersymmetric Models
}

\author{Antonio Delgado}

\address{ 
Dept. Physics and Astronomy, Johns Hopkins University \\
3400 N Charles Street, Baltimore MD 21218, USA.
\\ {\rm E-mail: adelgado@pha.jhu.edu}}

\abstract{
The minimal supersymmetric standard model, and extensions, predicts
a relatively light higgs particle if one supposes perturbativity until
high scales. That fact is in conflict with nowadays data coming from
LEPII fruitless
searches for the Higgs boson. This bound is sufficient to rule out
a big portion of the parameter space of the MSSM. In this letter we
propose two mechanism of raising the higgs mass at tree level, thus
not needing very heavy sparticles that would reintroduce
some fine-tuning in the problem.}

\normalsize\baselineskip=15pt

\section{Introduction}
The minimal supersymmetric standard model (MSSM) has been the most
promising candidate for physics beyond the standard model (SM)
for the last two decades. It addresses one of the biggest
problems in the SM, the hierarchy problem, the quadratic
sensitivity of the higgs mass to high scales, while giving
small contributions to the EW observables. At the same
time it gives a natural scenario for unification and has a candidate
for dark matter.

But the MSSM parameter space starts to be squeezed from the 
\emph{not-yet-found} higgs boson. The higgs mass in the MSSM
is predicted at tree-level to be smaller than $m_Z$, and this would
have completely ruled it out had not been for radiative corrections. The 
upper bound on the higgs mass in the MSSM is around 130 GeV
\cite{Carena:1995wu}, and this
suppose sparticles at 1 TeV and maximal mixing in the stop sector. These
numbers start to reintroduce fine tuning in the problem, since 
the cancellation in order to get the correct value for the EW
scale are on the order of few percent. 

One way to alleviate that problem is achieved in the next to minimal
supersymmetric standard model (NMSSM) where a new singlet coupled to the higgs
is added. This singlet can have two roles, one
is to solve the $\mu$ problem, the coincidence of the supersymmetric mass
of the higgses with the EW scale, the other one is to
 raise the tree level value of the 
higgs mass. With nowadays bounds the first case tends to produce the same
fine tuning as the MSSM, in the second case the bound on
the higgs mass is lifted to 150 GeV \cite{Espinosa:1998re}
 with the same assumptions as before. The gain is not too much
because perturbativity up to $M_{GUT}$ impose the actual
coupling of the singlet with the higgs to be small in the IR since
it is \emph{not} asymptotically free.

In this letter we will present two alternatives to increase
the tree level value, both based on the same model, 
making use of \emph{asymptotically free} interactions
thus letting higher values of coupling in the IR. This letter
is based on these two papers \cite{Batra:2003nj} and \cite{Batra:2004vc},
the reader should refer to those for further details and references.

\section{D-Term (gauge extension)}

The model we study is a gauge extension of the MSSM with gauge group
$SU(3)_c\times SU(2)_1\times SU(2)_2\times U(1)_y$, the $SU(2)$
structure as follows, the third family plus the usual
higgses are charged under the first $SU(2)$ whereas the first two families
are charged under the second one, with this assignments
the first $SU(2)$ is asymptotically free. Additional higgslike field exists
on the second $SU(2)$ to generate appropriate yukawa interactions. To break 
the $SU(2)_1 \times SU(2)_2$ to the diagonal
subgroup we use an extra bi-doublet $\Sigma$ which transforms as a
$(2, \overline{2})$.   Above the scale of diagonal symmetry breaking,
the $SU(2)_1 \times SU(2)_2$ $D$-term is
\begin{equation}
\frac{g_1^2}{8} \left(
{\rm Tr} \left[ \Sigma^\dagger \sigma^a \Sigma \right]
+ H^{\dagger}\sigma^a H -\overline{H} \sigma^a \overline{H}^{\dagger} +
\ldots \right)^2
+\frac{g_2^2}{8} \left(
{\rm Tr} \left[ \Sigma \sigma^a \Sigma^\dagger \right] \right)^2 .
\label{eq:d-term}
\end{equation}
The superpotential $\mathcal{W}=\lambda  S \left( \frac{1}{2}\Sigma\Sigma + w^2
\right)$
with an additional soft-mass $m^2$ for $\Sigma$ leads to the scalar potential
\begin{eqnarray}
  V_\Sigma & = & \frac{1}{2}B \Sigma \Sigma + h.c. + m^2 |\Sigma|^2 +
\frac{\lambda^2}{4} |\Sigma \Sigma|^2.
\label{eq:potential}
\end{eqnarray}
Here, $\Sigma \Sigma$ is contracted with two epsilon tensors
and $B=\lambda w^2$.
For large $B$, $\Sigma$ acquires a VEV,
$\langle \Sigma \rangle = u \mathbf{1}$, with
$u^2=(B - m^2)/\lambda^2$,
which breaks $SU(2)_1 \times SU(2)_2$ to the diagonal subgroup. The minimum
lies in a $D$-flat direction, leaving both Higgs fields massless.
                                                                               
Under the remaining $SU(2)$, $\Sigma$ contains a complex triplet, $T$,
along with a complex singlet.  Integrating out the real part of the heavy
triplet at tree-level gives
the effective Higgs potential below the triplet mass,
\begin{eqnarray}
&& \frac{g^2}{8} \:\Delta\: \left(H^{\dagger}
\vec{\sigma} H- \overline{H} \vec{\sigma} \overline{H}^\dagger  \right)^2
+ \frac{g_Y^2}{8} \left( |\overline{H}|^2-|H|^2 \right)^2,
 \nonumber \\
&&{\rm with\ \ }
\Delta=\frac{1+\frac{2 m^2}{ u^2}\frac{1}{g_2^2}}{1+\frac{2 m^2}{ u^2}
\frac{1}{g_1^2+g_2^2}}\hspace{.1in}  {\rm and \ \ }
\frac{1}{g^2} = \frac{1}{g_1^2} + \frac{1}{g_2^2}.
\label{eq:delta}
\end{eqnarray}

The MSSM $D$-term is recovered in the limit $u^2 \gg m^2$ (no SUSY breaking),
for which SUSY protects the $D$-term below the gauge-breaking scale.

Electroweak symmetry breaking occurs under the same conditions as in the MSSM.
 We find
the tree-level $W$ and $Z$ masses are corrected by the same relative
amount, $(1 - g^4 v^2 / 2 g_2^4 u^2 + ...)$ while the tree-level Higgs mass
satisfies
\begin{equation}
m_{h^o}^2 <  \frac{1}{2} \left( g^2 \Delta + g_Y^2 \right) v^2
\cos^2{2 \beta} .
\end{equation}
To maximize the upper bound, $\Delta$ should be made as
large as possible by sending $g_1 \rightarrow \infty$,
$g_2 \rightarrow g$ and $m^2 \gg u^2$ by as much as possible without
introducing fine-tuning. Also we have to be consistent with the EW fit.

We take the following example parameters:
\begin{itemize}
  \item $g_1(u) = 1.80, \ g_2(u) = .70$, inspired by a GUT
        with $g_1(\Lambda_{GUT}) = .97$. Additional spectator fields
 are 
included in the running to aid in unification.
  \item $u = 2.4$ TeV, above the lower limit from electroweak
        constraints, giving $M_{W'}, M_{Z'} \sim 4.5$ TeV.
\item $m=10\  {\rm TeV}$. One-loop finite corrections to the Higgs
mass parameter from supersymmetry breaking are $< 300$ GeV
whereas two-loop RGE contributions can be somewhat
larger if one assumes high-scale supersymmetry breaking.
\end{itemize}
We find $\Delta =  6.97$ and $m_h =214$ GeV
at tree-level in the large $\tan{\beta}$ and decoupling limits.
Loop corrections to the effective potential from the
top sector and the additional physics will make a
relatively small shift in the tree-level result. This
higgs mass is much greater that the MSSM or the NMSSM bound
thus allowing for much larger parameter space and less fine-tuning.

\section{F-term (NMSSM)}

The second proposal to increase the higgs mass is similar
to the usual NMSSM, we add an extra singlet with the following
coupling in the superpotential to the higgs fields,
$W=\lambda S H \overline{H}$, but with the same
gauge structure of the previous section. In the original NMSSM,
i.e. without gauge extension, the maximum value of $\lambda$ is relatively
small if one supposes perturbativity to the GUT scale. The RGE of $\lambda$ 
can be written as:

\begin{equation}
\frac{d\lambda}{dt}=\frac{\lambda}{16\pi^2}(3 |y_t|^2+4|\lambda|^2-3 g^2)
\end{equation}

\noindent
where $y_t$ is the top yukawa and $g$ is the $SU(2)$ coupling. Since $g$ has
a small value, the RGE for $\lambda$ is dominated by the first two terms and
these make $\lambda$ non asymptotically free and hence pretty small in the 
IR. However in our framework $g$ should be replaced by the strong coupling 
$g_1$ and in this case the RGE evolution for $\lambda$ allows for much larger
values in the IR.

The  one-loop Higgs CP-even mass matrix can be written as
a function of the CP-odd mass ($m_A$), the $\mu$ parameter, the stop
masses $m_{\tilde{t_i}}$, and the stop mixing parameter, $A_t$:

\begin{eqnarray}
 m_{11}^2&=&m_Z^2
\cos^2\beta+m_A^2\sin^2\beta-\frac{3 y_t^2}
{16\pi^2}\mu^2\frac{Z^2}{3}\nonumber\\ m_{22}^2&=&m_Z^2
\sin^2\beta+m_A^2\cos^2\beta+\frac{3 y_t^2} {16\pi^2}\left(4 m_t^2
\log\frac{m_{\tilde{t_1}} m_{\tilde{t_2}}}{m_t^2}+A_t(2 m_t Z-
A_t\frac{Z^2}{3})\right) \nonumber\\ m_{12}^2&=&-\frac{1}{2}
(m_Z^2+m_A^2-2v^2\lambda^2)\sin 2 \beta+\frac{3 y_t^2} {16\pi^2}\mu
\left(m_t Z-A_t\frac{Z^2}{3}\right)
\label{one-loop}
\end{eqnarray}
where:
\begin{equation}
Z=\frac{m_t(A_t+\mu\cot\beta)}{m_t^2+\frac{1}{2}(m^2_Q+m^2_U)}
\end{equation}
and the stop masses are defined with respect 
to the soft-masses for left- and right-handed stops ($m_Q$,$m_U$) as:
\begin{eqnarray}
m^2_{\tilde{t}_{1,2}}&=&m_t^2+\frac{1}{2}(m^2_Q+m^2_U)\pm W\nonumber\\
W^2&=&\frac{1}{4}(m^2_Q-m^2_U)^2+y_t^2 v^2 |A_t \sin\beta-\mu \cos\beta|^2
\end{eqnarray}

The charged Higgs mass is (at one-loop):
\begin{equation}
m^2_{H^{\pm}}= m^2_{A}+m^2_W - v^2 \left|\lambda\right|^2.
\end{equation}

From the above formulae there are several consequences that can be drawn. In
the decoupling limit ($m_A\to\infty$) and $\tan \beta \sim 1$ the lightest
higgs mass is $\lambda v$ which can be as large as 214 $GeV$. In contrast
to the usual MSSM and NMSSM where $\tan\beta$ has to be greater than 1.8 due 
to the perturbativity of $y_t$, in our case much lower values of $\tan\beta$
are allowed since $g_1$ also affects the RGE for $y_t$. There are regions
of the parameter space where the lightest of the higgs bosons is the
charged higgs giving rise to interesting new signatures in colliders since
the branching ratios are very different to those of the MSSM. 
Finally the LEPII 
bound of 114 $GeV$ can be avoided for small values of $m_A$ because in
that case the scalar that couples to $W$ and $Z$ bosons is the heaviest hence
that bound should apply to that higgs and not to the lightest.

\section{Conclusions}

The LEPII bound on the higgs mass makes the MSSM a fine-tuned theory. In this 
letter the bound is avoided thanks to a new strong interaction in the
gauge sector broken at a similar scale as SUSY. 
In one case a non-decoupling D-term  and in the second case a new 
superpotential interaction (NMSSM like) with large IR value 
give a new contribution to the higgs tree-level mass and large sparticle
masses are not needed to accomplish the experimental bound. Both
scenarios have interesting collider signatures.

\bibliographystyle{plain}

\begin{thebibliography}{99}


\bibitem{Carena:1995wu}
Y.~Okada, M.~Yamaguchi and T.~Yanagida,
Prog.\ Theor.\ Phys.\  {\bf 85}, 1 (1991);
M.~Carena, M.~Quiros, and C.~E.~M.~Wagner,
Nucl.\ Phys.\ B {\bf 461}, 407 (1996)
[arXiv:hep-ph/9508343];
H.~E.~Haber, R.~Hempfling, and A.~H.~Hoang,
Z.\ Phys.\ C {\bf 75}, 539 (1997)
[arXiv:hep-ph/9609331];
J.~R.~Espinosa and I.~Navarro,
Nucl.\ Phys.\ B {\bf 615}, 82 (2001)
[arXiv:hep-ph/0104047].
K.~Tobe and J.~D.~Wells,
Phys.\ Rev.\ D {\bf 66}, 013010 (2002)
[arXiv:hep-ph/0204196].
%
\bibitem{Espinosa:1998re}
J.~R.~Espinosa and M.~Quiros,
Phys.\ Rev.\ Lett.\  {\bf 81}, 516 (1998)
[arXiv:hep-ph/9804235].
%
\bibitem{Batra:2003nj}
P.~Batra, A.~Delgado, D.~E.~Kaplan and T.~M.~P.~Tait,
JHEP {\bf 0402}, 043 (2004)
[arXiv:hep-ph/0309149].
%
\bibitem{Batra:2004vc}
P.~Batra, A.~Delgado, D.~E.~Kaplan and T.~M.~P.~Tait,
JHEP {\bf 0406}, 032 (2004)
[arXiv:hep-ph/0404251].
%
\end{thebibliography}

\end{document}